# Epitaxial growth and characterization of high quality $Bi_2O_2Se$ thin films on $SrTiO_3$ substrates by pulsed laser deposition

Yekai Song[1,2,3], Zhuojun Li[1,2,*], Hui Li[4], Shujie Tang[1,2], Gang Mu[1,2], Lixuan Xu[1,2,5], Wei Peng[1,2], Dawei Shen[1,2], Yulin Chen[3], Xiaoming Xie[1,2,3] and Mianheng Jiang[1,2,3]

[1]State Key Laboratory of Functional Materials for Informatics, Shanghai Institute of Microsystem and Information Technology, Chinese Academy of Sciences, Shanghai 200050, China.

[2]CAS Center for Excellence in Superconducting Electronics (CENSE), Shanghai 200050, China

[3]School of Physical Science and Technology, ShanghaiTech University, Shanghai 200031, China

[4]College of Engineering Physics, Shenzhen Technology University, Shenzhen 518118, China

[5]University of Chinese Academy of Sciences, Beijing 100049, China

*E-mail: lizhuojun@mail.sim.ac.cn

**Abstract**

Recently, $Bi_2O_2Se$ is discovered as a promising two-dimensional (2D) semiconductor for next generation electronics, due to its moderate bandgap size, high electron mobility and pronounced ambient stability. Meanwhile, it has been predicted that high quality $Bi_2O_2Se$-related heterostructures may possess exotic physical phenomena, such as piezoelectricity and topological superconductivity. Herein, we report the first successful heteroepitaxial growth of $Bi_2O_2Se$ films on $SrTiO_3$ substrates via pulsed laser deposition (PLD) method. Films obtained under optimal conditions show an epitaxial growth with the *c* axis perpendicular to the film surface and the *a* and *b* axes parallel to the substrate. The growth mode transition to three dimensional (3D) island from quasi-2D layer of the heteroepitaxial $Bi_2O_2Se$ films on $SrTiO_3$ (001) substrates is observed as prolonging deposition time of films. The maximum value of electron mobility reaches 160 $cm^2/V^{-1}s^{-1}$ at room temperature in a 70nm-thick film. The thickness dependent mobility provides evidence that interface-scattering is likely to be the limiting factor for the relatively low electron mobility at low temperature, implying that the interface engineering as an effective method to tune the low temperature electron mobility. Our work suggests the epitaxial $Bi_2O_2Se$ films grown by PLD are promising for both fundamental study and practical applications.

**Introduction**

Two-dimensional (2D) layered $Bi_2O_2Se$ is provoking immense interests owing to its great potential for next generation electronics. As a novel 2D layered semiconductor, $Bi_2O_2Se$ was discovered to exhibit an ultrahigh electron mobility ($>2\times10^4$ $cm^2V^{-1}s^{-1}$) at low temperatures[1]. Scanning tunneling spectroscopy (STS) study reveals that $Bi_2O_2Se$ hosts an unusual spatial uniformity of the bandgap (~0.8 eV) without undesired in-gap states on the sample surface even with up to ~50% vacancy defects[2]. Moreover, heat and humidity treatment experiments demonstrate ultrathin $Bi_2O_2Se$ plates are excellent in environmental stability[1], which is of great importance in the device performance and reliability. Benefited from the above-mentioned advantages, $Bi_2O_2Se$ shows high performance in field effect transistors[1], photodetectors[3-5] and magnetoresistance devices[6]. On the other hand, theoretical investigations have predicted that $Bi_2O_2Se$ possesses piezoelectricity and ferroelectricity upon a certain in-plane strain[7], which could be achieved by epitaxial growth on the substrates with the in-plane lattice constant a few percent larger than that of $Bi_2O_2Se$. Of even greater interest, the heterostructures between $Bi_2O_2Se$ and copper oxides as well as iron-based superconductors are suggested to be candidates for exploring novel phenomena such as topological superconductivity[2, 8, 9]. Therefore, successful preparation of epitaxial films is important and prerequisite to explore both the exotic phenomena and potential applications in $Bi_2O_2Se$ and related heterostructures.

The chemical vapor deposition (CVD) method has been developed to fabricate high quality 2D $Bi_2O_2Se$ films on various kinds of substrates, such as the Mica, $SrTiO_3$, $LaAlO_3$ and MgO[10-12]. The ambient pressure vapor-solid growth approach for the preparation of 2D $Bi_2O_2Se$ nanosheets on the Mica is reported recently[13]. Pulsed laser deposition, as one of the typical physical vapor deposition (PVD) techniques, has been proven to be an alternative way to the CVD method to obtain the layered 2D materials such as few-layer graphene[14], $MoS_2$[15], $WS_2$[16] and Boron Nitride[17]. The merits of accurate replication of stoichiometry from bulk target to films, in-situ fabrication of heterostructure via sequentially switching different targets without breaking the vacuum make it potential in growing complex oxide film[18, 19], multi-heterostructure[20] and superlattice[21] by PLD method. Therefore, PLD technique is appealing for the growth of hetero-epitaxial $Bi_2O_2Se$ films.

In this paper, we report the first successful epitaxial growth of $Bi_2O_2Se$ films on $SrTiO_3$ (001) using PLD method. X-ray diffraction (XRD) characterizations demonstrate the appropriate temperature for high quality Bi$_2$O$_2$Se epitaxial growth is 425°C. The growth mode transition from the quasi-2D layer to 3D island is observed for the growth of Bi$_2$O$_2$Se on SrTiO$_3$ substrates by XRD and atomic force microscopy (AFM). Transmission electron microscopy (TEM) images reveal a sharp interface between Bi$_2$O$_2$Se films and SrTiO$_3$ substrates. The maximum of electron mobility reaches 160 cm$^2$V$^{-1}$s$^{-1}$ at room temperature in a 70nm-thick film. Phonon-scattering is the main mechanisms in determining the electron mobility in PLD-grown Bi$_2$O$_2$Se/SrTiO$_3$ heterostructures at high temperatures. Meanwhile, the interface-scattering is the limiting factor for the relatively low mobility at low temperatures indicated by the thickness-dependent mobility. Our results provide a new method to fabricate Bi$_2$O$_2$Se films and pave a way for exploring novel physical properties and potential applications in Bi$_2$O$_2$Se-related heterostructures.

**Results and discussion**

Figure 1(a) shows the tetragonal crystal structure of Bi$_2$O$_2$Se, which can be regarded as the alternative stacking of negatively charged planar Se layers and positively charged Bi$_2$O$_2$ layers along *c* direction. Thus the weak electrostatic interactions hold the layers. As shown in Figure 1(b), powder XRD patterns of home-made target can be well indexed to the crystal structure of Bi$_2$O$_2$Se without other impurity phases. The deduced lattice parameters *a* =3.886 Å and *c* = 12.202 Å are consistent with those published previously[22].

Single crystalline SrTiO$_3$ was chosen as the substrate in considering the small lattice mismatch (~0.6%) with respect to Bi$_2$O$_2$Se. Figure 1(c) shows the out-of-plane $2\theta-\theta$ patterns of Bi$_2$O$_2$Se thin films on SrTiO$_3$ substrates under various growth substrate temperature $T_s$. Films are not in crystallization under $T_s$=300°C. Increasing $T_s$ to 400°C leads to the formation of Bi$_2$O$_2$Se phase with the out-of-plane orientation along *c*-axis. Small amount of impurity Bi$_8$Se$_7$ is detected as well, which is marked as asterisks (*) in the Figure 1(c). At higher $T_s$ of 425~450°C, only (00*l*) diffraction peaks from Bi$_2$O$_2$Se can be observed without any secondary phases. With increasing of $T_s$ from 450°C to 500°C, the preferred out-of-plane orientation of Bi$_2$O$_2$Se films would evolve from (002) to (013). At the highest $T_s$=550°C, no crystalline phases are found to form on substrates. For clarity, the effects of substrate temperature on crystallization and orientation of Bi$_2$O$_2$Se films are summarized in Figure 1(d). In the present work we mainly focus on the Bi$_2$O$_2$Se films with (00*l*) preferred orientation.

We further checked the crystallinity of two 23nm-thick $Bi_2O_2Se$ samples deposited at 425°C and 450°C. As shown in Figure 2(a), the (004) diffraction peak of the films prepared at 425°C displays clearer Laue oscillation fringes, suggesting that the out-of-plane order is high and coherent over the entire film thickness. Accordingly, for $T_s$=425°C, the full width at half-maximum (FWHM) of the $Bi_2O_2Se$ (004) peak in rocking curve was as narrow as 0.08° (the inset of Figure 2(b)), indicating a high degree of crystallinity with low mosaicity. The FWHM value increases to 0.13° in 23 nm-thick $Bi_2O_2Se$ film deposited at 450°C, suggesting the relatively poor crystallinity. In order to reveal the in-plane texture, the azimuthal ϕ scans of the off-axis {011} peaks of the 23 nm-thick $Bi_2O_2Se$ films deposited at 425°C were further performed. As shown in Figure 2(c), four $Bi_2O_2Se$ {011} peaks are spaced by 90 degrees and four STO {011} peaks with the same spacing are also found at the same angular positions, suggesting in-plane epitaxial growth. Hence, the optimized substrate temperature of 425°C is determined and the $Bi_2O_2Se$ films discussed hereafter in the context are those prepared at $T_s$=425°C. Figure 2(d) shows the (004) diffraction peaks of several $Bi_2O_2Se$ films with different thicknesses. With the increasing of $Bi_2O_2Se$ thickness, the Laue oscillation fringes around the (004) reflections gradually emerge. However, the fringes are not clear in the ~50 nm-thick $Bi_2O_2Se$ film (green line) and become undiscernible in ~60 nm-thick sample (blue line). The evolvement of the Laue oscillation in Figure 2(d) implies a progressive relaxation of $Bi_2O_2Se$ films with film thicknesses increasing. As is shown in Figure 2(e), the lattice parameter $c$ derived from (004) diffraction peak decreases gradually with the increasing of films thicknesses, which is probably due to the increasing of Se vacancies in the as-grown $Bi_2O_2Se$ films generated during the deposition. The variation of films thickness with deposition time shown in Figure 2f indicates the growth rate is ~ 6 nm/min for $Bi_2O_2Se$ grown on $SrTiO_3$, which is about 10 times faster than that of the CVD method[12]. The relatively fast growth rate makes it possible to produce high-quality $Bi_2O_2Se$-based semiconductor devices efficiently in the future.

To further explore the nucleation and growth evolution, surface morphology of the $Bi_2O_2Se$ films at different growth stages was characterized by ex-situ AFM. Figure 3(a)-(d) represent the AFM images of four $Bi_2O_2Se$ films deposited with 2~10 minutes. At the beginning of the growth, the $Bi_2O_2Se$ spices nucleate at the substrate surface and grow up in the form of square-like shape 2D islands marked by yellow right angle in Figure 3(a). The lateral size of the 2D island is several tens of nanometers and the thickness is 5~6 nm, according to the height profile in Figure 3(e). The square-like 2D islands also align along the same direction. As shown in Figure 3(b)-(c), with the increase of the deposition time, the 2D islands grow up and gradually coalesce into uniform and continuous $Bi_2O_2Se$ thin films on the $SrTiO_3$ substrate. The $Bi_2O_2Se$ film deposited with 6 minutes shows a relatively smooth surface with a root-mean square (RMS) surface roughness of ~0.93 nm. Further increasing of the deposition time to 10 minutes results in the formation of 3D islands on the film surface, leading to a rather large RMS surface roughness of ~5.12 nm. Based on the above results, the schematic diagrams of the entire growth processes are depicted in Figure 3(i)-(m). Growth mode transition from the quasi-2D layer to 3D island growth can therefore be proposed in our $Bi_2O_2Se$ films deposited on $SrTiO_3$ substrate by PLD. In the case of heteroepitaxial growth, the thermodynamic factors, e.g. lattice misfit, are important for growth mode[23]. The lattice mismatch between $Bi_2O_2Se$ and $SrTiO_3$ is considerably small (~0.6%), implying that under proper growth conditions one should get very smooth $Bi_2O_2Se$ films, which would give clear Laue oscillation fringes around the Bragg peak, as indicated in XRD patterns in Figure 2(d) (films with thickness less than 50 nm). In addition, the morphology of the 2D island is affected by both thermodynamic and kinetic factors, e.g. the crystal lattice symmetry of the substrates surface, the diffusion rate of as-deposited adatoms on the substrate surface as well as at the step edges of 2D islands[23]. Hence it is reasonable that the formation of square-like $Bi_2O_2Se$ islands on the $SrTiO_3$ (001) substrates in a growth regime (temperature, pressure). Similar phenomenon has been observed in the growth of $Bi_2O_2Se/SrTiO_3$ using CVD method[12]. As the growth proceeds, the elastic strain induced by the misfit between the films and substrate is becoming large and would be relaxed accompanied by the formation of 3D islands on the film surface. The absence of Laue oscillation fringes in our $Bi_2O_2Se$ films thicker than 50nm reflects the change of film surface roughening (Figure 2(d)). The characterization of surface morphology of the $Bi_2O_2Se$ films with different thicknesses is consistent with Laue oscillation observed by XRD in $\theta$-$2\theta$ scan. Scanning transmission electron microscopy (STEM) combined with energy dispersive X-ray spectroscopy (EDX) mapping was used to examine the micro structure of the optimized samples deposited at 425°C. As shown in Figure 4(a), the thickness of the film measured from the cross sectional image is ~23 nm, consistent with the result deduced from XRD data. With a higher magnification, the atomically sharp interface between $Bi_2O_2Se$ and $SrTiO_3$ are observed in Figure 4(b). A well-defined space of ~0.60 nm obtained along the out of plane direction matches well with the monolayer thickness of $Bi_2O_2Se$ (0.61 nm). Figure 4(c) is the expanded view of the high-angle annular dark-field STEM (HAADF-STEM) image from the square area marked in Figure 3(b), which clearly reveals the typical layered structure viewed along [001] direction of $Bi_2O_2Se$. Atomic resolution EDX mapping gives clear elemental analysis of the distribution of Bi, O, Se along out of plane direction. As shown in Figure 4(d)-(f), the alternate stacking of $Bi_2O_2$ layer and Se layer agrees well with the structure model of $Bi_2O_2Se$ as depicted in Figure 1(a).

Electrical transport measurements were performed on the $Bi_2O_2Se$ films with different thicknesses. Figure 5(a) shows temperature dependent sheet resistance ($R_S$-$T$) curves of the films with thicknesses ranging from ~27 nm to 83 nm. The films thinner than ~27 nm show insulating behaviors, which may results from the poor continuity in the initial growth stage as discussed above. Besides, interface induced electron scattering cannot be ruled out, which would be discussed later. The $Bi_2O_2Se$/$SrTiO_3$ films show metallic conducting behavior as the thickness increases. Hall resistance $R_{xy}$ as a function of the applied magnetic field B ($R_{xy}$-$B$ curves) measured at 300K and 2K are shown in Figure 4(b). The $R_{xy}$ of all the $Bi_2O_2Se$ films with different thicknesses show linear dependence with $B$ and the slopes of $R_{xy}$−$B$ curves are negative, indicating the electron carrier conducting behavior. The slope of the $R_{xy}$−$B$ curve decreases as the thickness increases, suggesting a much higher electron concentration in the thicker samples. To make a quantitative comparison, 2D carrier concentration ($n$) and Hall mobility ($\mu_H$) for the $Bi_2O_2Se$/$SrTiO_3$ films with different thicknesses are calculated and summarized in Figure 4(c)-(d). The $n$ and $\mu_H$ can be inferred from the equations: $n = -\frac{1}{eR_H}$ and $\mu_H = \frac{1}{R_S n e}$, respectively. $R_H$ is the Hall coefficient, which is defined as the slope of the $R_{xy}$−B curve and $e$ is the charge of an electron. As the thickness of the $Bi_2O_2Se$ film increases, $\mu_H$ at room temperature increases first, then approaches 120 $cm^2V^{-1}s^{-1}$ for the 40 nm-thick film, and finally maintains the constant value of 120 $cm^2V^{-1}s^{-1}$ for thicker samples. The maximum mobility is 160 $cm^2V^{-1}s^{-1}$ in a 70nm-thick $Bi_2O_2Se$ film. Meanwhile $n$ increases monotonically with increasing the thickness of the $Bi_2O_2Se$ films. The evolution of $\mu_H$ and $n$ with thickness variation at 2K is similar to that at room temperature, as shown in Figure 4(d).

Prior to discussing the scattering mechanism in our PLD-grown $Bi_2O_2Se$ films, the origin of n-type conducting and electron carrier density should be discussed. Theoretically, there are three kinds of lattice defects acting as donors in $Bi_2O_2Se$, including Se vacancies ($V_{Se}$), O vacancies ($V_O$) and Bi antisites at Se position ($Bi_{Se}$). Due to the low formation energy of $V_{Se}$, loss of Se element usually occurs and has been found inducing electron doping effect in some selenides compounds[24, 25]. The high vacuum conditions during the deposition process also gives rise to the generation of $V_{Se}$ due to the high volatility of selenium. Considering that no extra selenium was added purposely in $Bi_2O_2Se$ target, the growth process can be regarded as Se-poor condition, under which $Bi_{Se}$ can unavoidably form as well. Oxygen vacancies are common defects during oxide film growth process[18, 26], which probably contribute to the electron doping in the $Bi_2O_2Se$ films. Previous theoretical calculation reveals that the formation energy of $V_O$ is much higher than those of $V_{Se}$ and $Bi_{Se}$ under Se-poor condition[27, 28]. Therefore, $V_{Se}$ and $Bi_{Se}$ are thought to be the main donor defects in our PLD-grown $Bi_2O_2Se$ films. As the growth of the films proceeds, it is reasonable that the $V_{Se}$ and $Bi_{Se}$ would accumulate and result in an increase of electron carrier concentration. The thickness-dependent carrier concentration in PLD-grown $Bi_2O_2Se$ varies in the order of $10^{13}$-$10^{14}$ cm$^{-2}$, comparable to the results reported previously[10, 12]. It is believed that defects tuning in PLD-grown $Bi_2O_2Se/SrTiO_3$ films can be realized through fine tuning of the PLD parameters and proper target preparation. Defect engineering would be helpful to further elucidate the origin of the carrier in $Bi_2O_2Se/SrTiO_3$ films in future. Based upon the foregoing discussion on the formation of electron carrier, next we try to understand the evolution of $\mu_H$ as a function of sample thickness through analyzing the scattering mechanism in PLD-grown $Bi_2O_2Se$ films. The temperature dependence of $\mu_H$ illustrated in Figure S3 demonstrates that phonon-scattering is the main mechanism in our $Bi_2O_2Se$ films at high temperatures, in accordance with the reported bulk single crystal[22] and CVD-grown 2D films[1]. The thickness dependence of the carrier density and mobility at high temperatures in Figure 5c shows that both the $\mu_H$ and $n$ increase as film thickness increases. Considering that the carrier density is an indication of the Se vacancy level, therefore, the $V_{Se}$ has a minor contribution to the electron scattering. Such a behavior agrees well to the proposed self-modulation-doping mechanism[27], in which, the conduction layer and scattering center are separated spatially. The saturation of the sheet resistance at low temperature (Figure 5a) indicates that the charged defects dominate the carrier scattering at low temperatures. Two distinct mobility evolution as function of thickness and carrier density are observed in Figure 5c and d, which corresponds to different scattering mechanism respectively. The high-temperature mobility is thickness and carrier density independent above a critical value ~40 nm which is expected for the phonon dominated scattering; while in the charged impurity dominated low-temperature case, the mobility increases as the carrier density increases and builds up the screening[29]. Besides, an anomaly of drastically drop of the carrier mobility (Figure 5c) and even insulating behavior at room temperature as thickness decrease is observed, indicating the scattering from interface charged impurity start to overwhelm the phonon scattering and finally lead to strong localization. Another anomaly is the saturation and even the drop of the mobility as the thickness reach a critical value ~70 nm at low temperature, which might be caused by the growth mode changing induced charged impurity increase.

Finally, a comprehensive comparison of electrical properties among our PLD-grown $Bi_2O_2Se$ films and other samples fabricated by different methods are summarized in table I. It can be found that the difference of mobility at 300K is relatively small for all the samples, even though there is two orders of magnitude discrepancy among their carrier densities. This is attributed to the phonon-dominating scattering at room temperature. In contrast, the mobility at 2K varies greatly from the order of $10^2$ $cm^2V^{-1}s^{-1}$ to $10^5$ $cm^2V^{-1}s^{-1}$. Worth noticing that the samples with lower mobility at 2K also show lower value of residual resistivity ratio (RRR=$R_{S,\ 300K}/R_{S,\ 2K}$), which probably due to the defects in crystal lattice and disordered atom arrangement near interface as discussed above. Compared with CVD-grown thin $Bi_2O_2Se$ flakes on Mica substrates by Peng's group[10], the rather low mobility at 2K in our PLD-grown $Bi_2O_2Se$ films also strengthen these inferences arising from the use of different substrates and the nature of the different growth methods. However, it is promising to realize further improvement of electrical performance via interface and defects engineering in PLD-grown $Bi_2O_2Se$ thin films.

**Conclusion**

In summary, we succeeded in growing high quality epitaxial $Bi_2O_2Se$ films on $SrTiO_3$ substrates by PLD method. The evolvement of surface morphology with deposition time demonstrates the growth mode transition from quasi-2D layer to 3D island of heteroepitaxial $Bi_2O_2Se$ films on $SrTiO_3$ (001) substrates by PLD method. Transport measurement results show phonon-limited electron carrier transport behaviors for the $Bi_2O_2Se/SrTiO_3$ heterostructures at room temperature. The maximum electron mobility reaches 160 $cm^2/V^{-1}s^{-1}$ at room temperature. Thickness dependent electrical transport properties suggest that charge impurity is the limitation factor for the low temperature mobility, especially for the very thin film, the interface defects play a destructive role to the conductivity. Our work indicates that interface engineering of the heterostructure films, such as pretreatment of the substrate surface or adding buffer layers, would be necessary for the application of the $Bi_2O_2Se$ film. Epitaxial $Bi_2O_2Se$ films on $SrTiO_3$ by PLD method provides a promising platform for exploring of exotic physical phenomena and the potential device applications.

**Materials and methods**

*Target fabrication*

Polycrystalline $Bi_2O_2Se$ targets were prepared by a two-step solid state reaction method. Firstly, high-purity starting materials $Bi_2O_3$ (4N), Se (4N) and Bi (4N) powders were loaded in $Al_2O_3$ crucibles, which were evacuated and sealed in quartz tubes. The tubes were heated to 950°C and kept for 24hrs to obtain the precursors. Secondly, to obtain high density $Bi_2O_2Se$ targets, the as-sintered precursors were ground thoroughly in an argon-filled glovebox, followed by heat treatment for 12hrs in a hot-pressing (450°C /6MPa) furnace. During this process, argon was continuously introduced to prevent the unfavorable reaction between the target materials and the atmosphere. Finally, the as-grown targets were checked by a powder x-ray diffractometer (DX-2700) and no trace impurities were detected.

*Thin film deposition*

The $Bi_2O_2Se$ thin films were grown on $SrTiO_3$ (001) single crystal substrates in a high vacuum ($\sim10^{-5}$Pa) chamber by PLD. A KrF excimer laser (Coherent, COMPexPro201) was employed to produce a 248-nm-wavelength laser beam. The energy density, repetition rate and target-substrate distance were 150mJ/mm$^2$, 4 Hz and 50mm, respectively. The substrate temperature varied from 300°C to 550°C in order to find the optimized deposition conditions. To facilitate the electrical transport measurements, the films were patterned to Hall bar configuration (3mm×0.5mm) using a metal mask during growth.

*Structure and morphology characterization*

The crystallinity and the lattice parameters of the as-grown films were evaluated by X-ray diffraction (XRD, Bruker, D8 Discover) with Cu K$_\alpha$ radiation using the high resolution mode. Surface morphology was observed by atomic force microscopy (Bruker, Dimension ICON). A step profile tester (Bruker, Dektak-XT) was employed for a convenient thickness measurement of $Bi_2O_2Se$ films. The microstructure examinations of the films were performed by transmission electron microscopy (JEOL,

JEM-ARM300F). High-angle annular dark-field scanning transmission electron microscopy (HAADF-STEM) using a double spherical aberration probe corrector was conducted to examine the atoms arrangement at the interface between $Bi_2O_2Se$ films and $SrTiO_3$ substrates. The energy-dispersive x-ray spectrometer (EDX) provided outstanding sensitivity to determine elements with atomic resolution.

*Electrical transport measurements*

The transport properties of the films were carried out using Physical Properties Measurement System (Quantum Design, Dynacool). The temperature dependent resistance was measured from 2K to 300 K and Hall measurements were taken with magnetic fields of up to 9T applied perpendicular to the film surface at 2K and 300K, respectively. Metallic indium was used to form favorable ohmic contacts (see Figure S2).

**Acknowledgments**

The authors thank Dr. Hua Jin and Dr. Lu Zhang for helpful discussions. XRD and AFM characterizations were supported by Superconducting Electronics Facility (SELF) in the Shanghai Institute of Microsystem and Information Technology. This work is financially supported by the National Natural Science Foundation of China (No. 11704395, No. 11227902), the Natural Science Foundation of Shanghai (No. 17ZR1436300), Young Innovative Talents Project for Regular Universities in Guangdong Province (No. 2018KQNCX396) and the "Strategic Priority Research Program (B)" of the Chinese Academy of Sciences (No. XDB04010600).

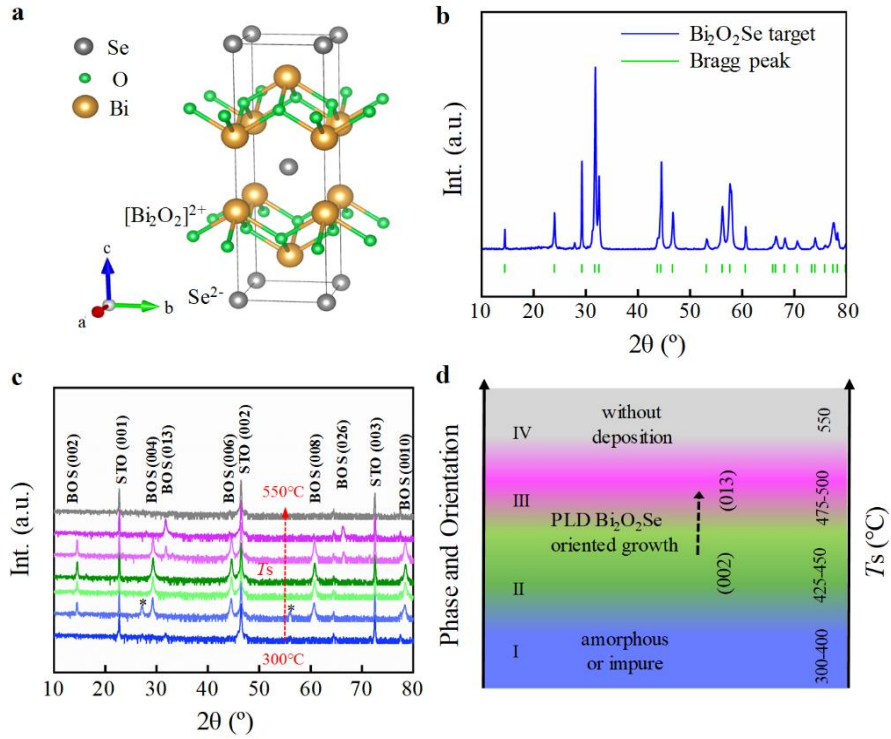

**Figure 1 (a)** Crystal structure of Bi$_2$O$_2$Se drawn based on the tetragonal structure (space group I4/mmm, No. 139, a=b=3.887 Å, c=12.164 Å). **(b)** X-ray diffraction pattern of the home-made polycrystalline Bi$_2$O$_2$Se target (blue line). Vertical green bars indicate the Braggs reflections for Bi$_2$O$_2$Se. **(c)** Out-of- plane 2$\theta$-$\theta$ XRD patterns for the Bi$_2$O$_2$Se films deposited on SrTiO$_3$ (001) substrates at different substrate temperate $T_s$. (*) marks the Bi$_8$Se$_7$ impurity phase. **(d)** A diagram of the substrate temperature $T_s$ dependent phase and out-of-plane orientation for PLD-grown Bi$_2$O$_2$Se films on SrTiO$_3$.

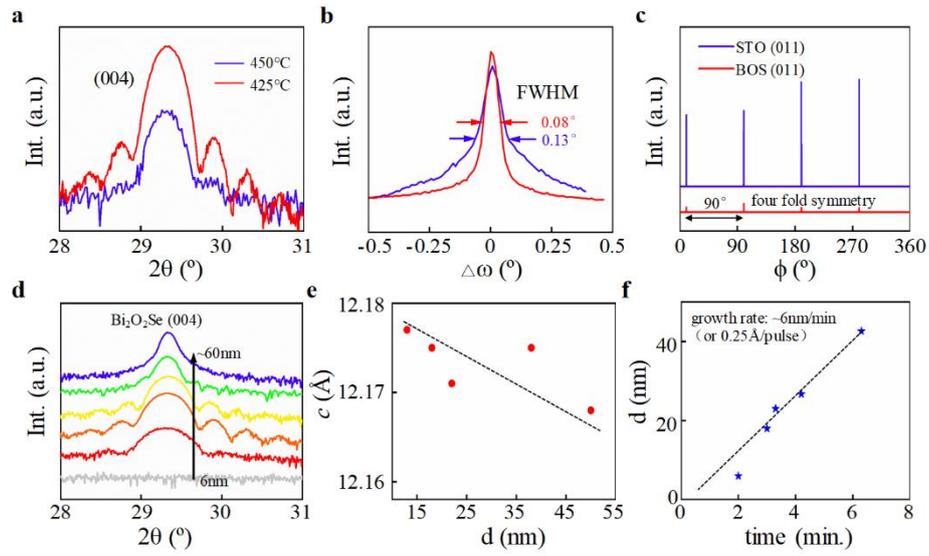

**Figure 2 (a)** Zoom in region of the (004) diffraction peak of two $Bi_2O_2Se$ films deposited at different $T_s$ (425°C, 450°C). The clear Laue oscillations indicate the smoothness of the film (uniform film thickness). **(b)** The rocking curves of corresponding (004) reflection. **(c)** Azimuth ϕ scans of the off-axis {011} peaks of $SrTiO_3$ substrate and the $Bi_2O_2Se$ film deposited at $T_s$ = 425°C, respectively. **(d)** X-ray diffraction patterns of the (004) diffraction peak of the $Bi_2O_2Se$ films with different thicknesses. **(e)** Variation of the lattice parameters $c$ of the $Bi_2O_2Se$ films with film thicknesses calculated from the (004) diffraction. **(f)** Growth rate plotted by deposition duration dependence of the $Bi_2O_2Se$ film thickness.

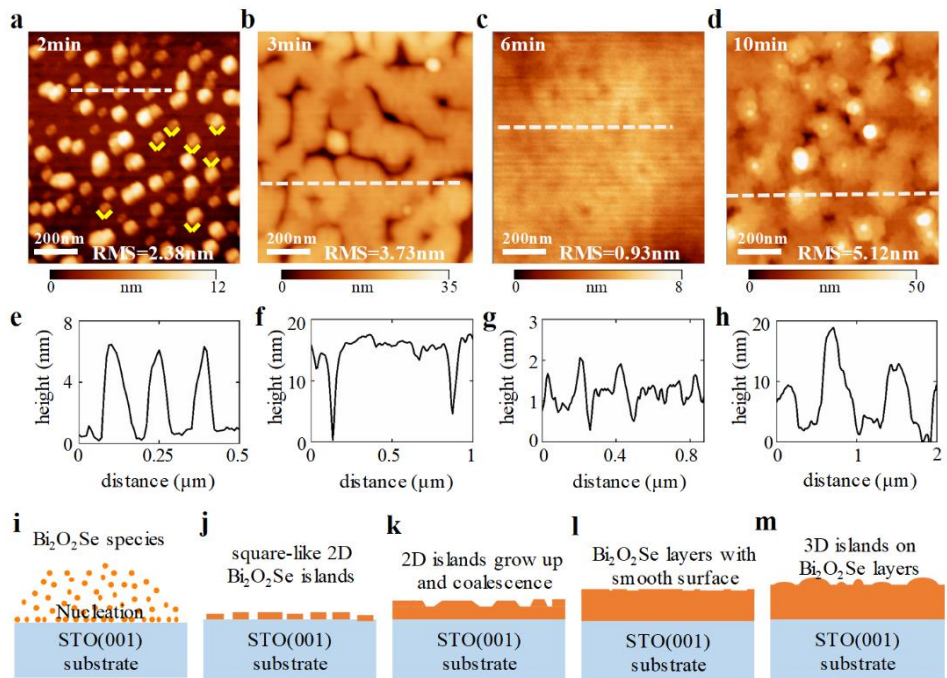

**Figure 3** Surface morphology and structure evolution of $Bi_2O_2Se$ films deposited on $SrTiO_3$ substrates with different deposition time. **(a)-(d)** AFM images of the $Bi_2O_2Se$ films at different stages of the growth. The yellow right angles in Figure (a) indicate the shape and orientation of the $Bi_2O_2Se$ domains. **(e)-(h)** Corresponding height profile of the white dashed line-marked scan in figure (a)-(d). **(i)-(m)** Schematic diagram of the growth mode for $Bi_2O_2Se$ films grown on $SrTiO_3(001)$ substrates by PLD method.

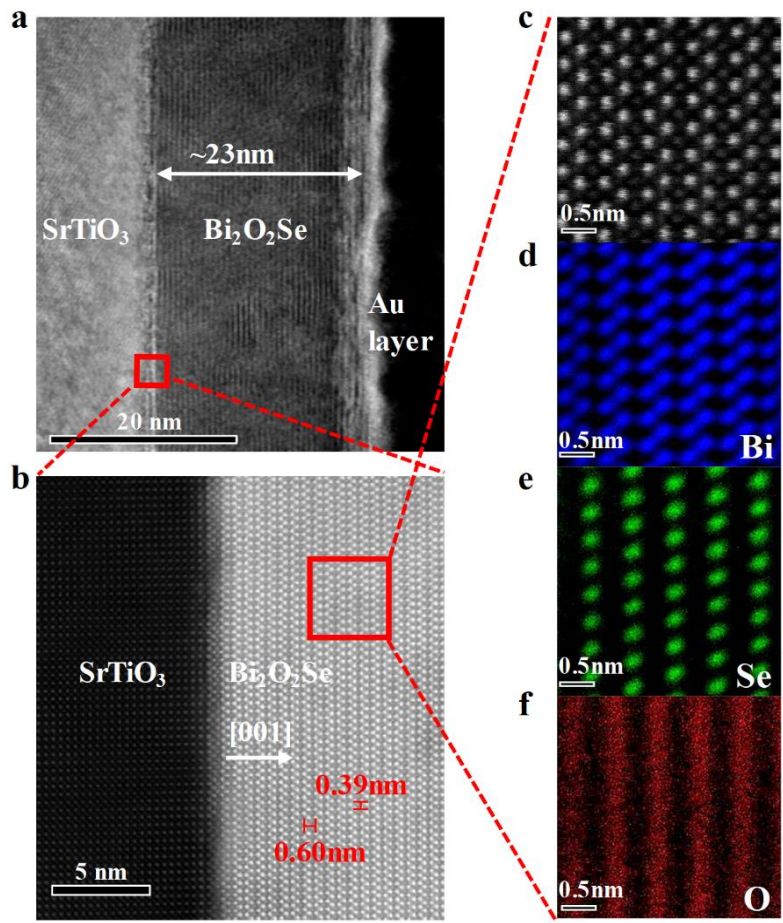

**Figure 4** Cross-section examination of the $Bi_2O_2Se/SrTiO_3$ film deposited at $T_s$=425°C by TEM characterizations. **(a)** Cross-sectional low-magnification TEM image of the $Bi_2O_2Se$ film. The film thickness measured from the image is ~23 nm, consistent with that determined from XRD data. **(b)** HAADF-STEM image with atomic resolution enlarged from the rectangular area in (a). **(c)** Corresponding HAADF-STEM image enlarged from the square area in (b). **(d)-(f)** Corresponding EDX mapping of Bi, O, Se elements distributed along the out of plane direction of the $Bi_2O_2Se/SrTiO_3$ film.

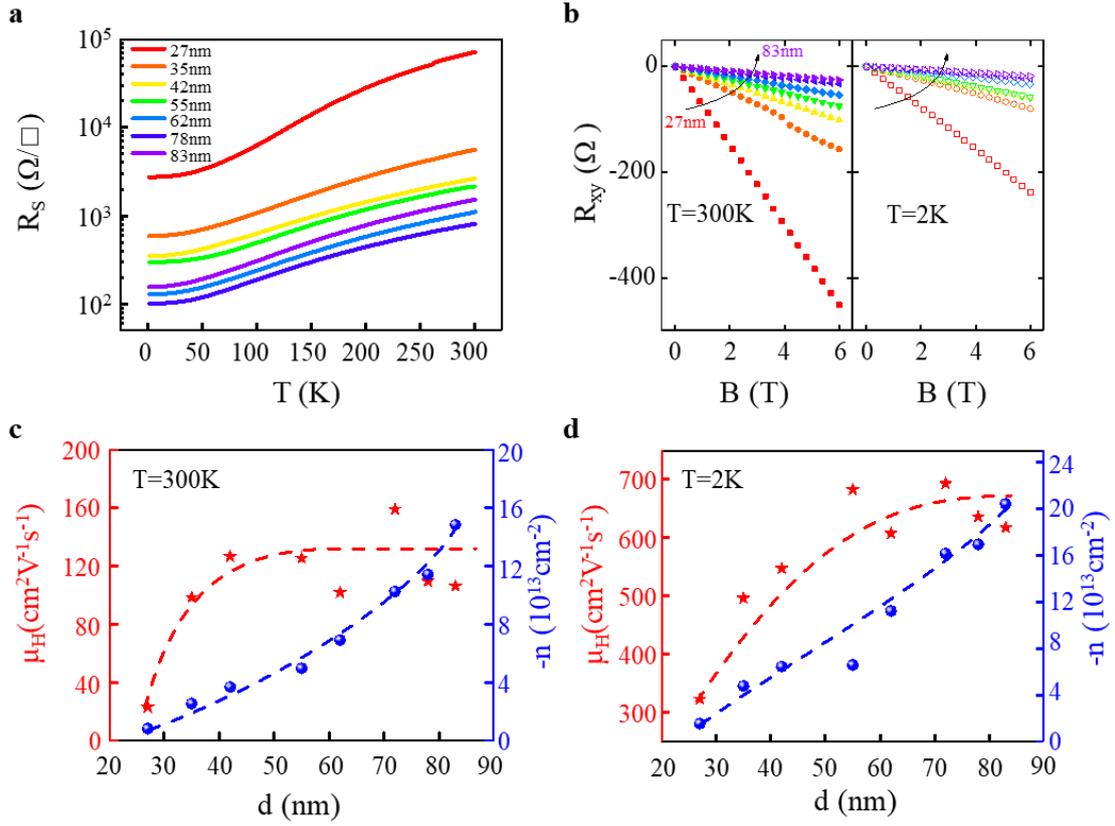

**Figure 5** Thickness dependent electrical transport properties for $Bi_2O_2Se$ films. **(a)** Sheet resistance versus Temperature ($R_s$-$T$) curves. **(b)** Hall resistance $R_{xy}$ versus the external magnetic field measured at 300K (left) and 2K (right). **(c)-(d)** Statistics and comparison of Hall mobility and carrier concentration as a function of the $Bi_2O_2Se$ film thickness at 300K (c) and 2K (d).

| Synthetic Method | Substrate | Thickness (nm) | Residual Resistivity Ratio | Carrier density (cm$^{-3}$) | Hall Mobility@300K (cm$^2$V$^{-1}$s$^{-1}$) | Hall Mobility@2K (cm$^2$V$^{-1}$s$^{-1}$) | Ref. |
|---|---|---|---|---|---|---|---|
| modified Bridgman | \ | bulk | 585 | $8\times10^{18}$ | 370 | $2.8\times10^5$ | Sci. Adv. 2018, 4, eaat8355 |
| gas-phase transportation | \ | bulk | ~40 | $5\times10^{18}$ | ~10 | ~700 | Journal of Crystal Growth, 2018, 498, 244 |
| CVD | Mica | 1L~6L | ~60 | $10^{18~19}$ | 313 | ~$2\times10^4$ | Nanolett. 2017, 17, 3021 |
| CVD | Mica | 3~9 | \ | $10^{16~18}$ | 150~450 | \ | Nanolett. 2019, 19, 197 |
| CVD | SrTiO$_3$ | 1~30 | ~6(10nm) | \ | 94(10nm) | \ | Nanolett. 2019, 19, 2148 |
| PLD | SrTiO$_3$ | 6~80 | 7~26 | $10^{18~19}$ | 15~160 | 100~700 | This work |

Table 1. Comparison of electrical performance of Bi$_2$O$_2$Se films prepared by different methods

# Supplementary Information

# Epitaxial growth and characterization of high quality $Bi_2O_2Se$ thin films on $SrTiO_3$ substrates by pulsed laser deposition

Yekai Song[1,2,3], Zhuojun Li[1,2,*], Hui Li[4], Shujie Tang[1,2], Gang Mu[1,2], Lixuan Xu[1,2,5], Wei Peng[1,2], Dawei Shen[1,2], Yulin Chen[3], Xiaoming Xie[1,2,3] and Mianheng Jiang[1,2,3]

[1]State Key Laboratory of Functional Materials for Informatics, Shanghai Institute of Microsystem and Information Technology, Chinese Academy of Sciences, Shanghai 200050, China.

[2]CAS Center for Excellence in Superconducting Electronics (CENSE), Shanghai 200050, China

[3]School of Physical Science and Technology, ShanghaiTech University, Shanghai 200031, China

[4]College of Engineering Physics, Shenzhen Technology University, Shenzhen 518118, China

[5]University of Chinese Academy of Sciences, Beijing 100049, China

*E-mail: lizhuojun@mail.sim.ac.cn

# Supplementary Text

## TEM characterization of as-synthesized $Bi_2O_2Se$ thin films

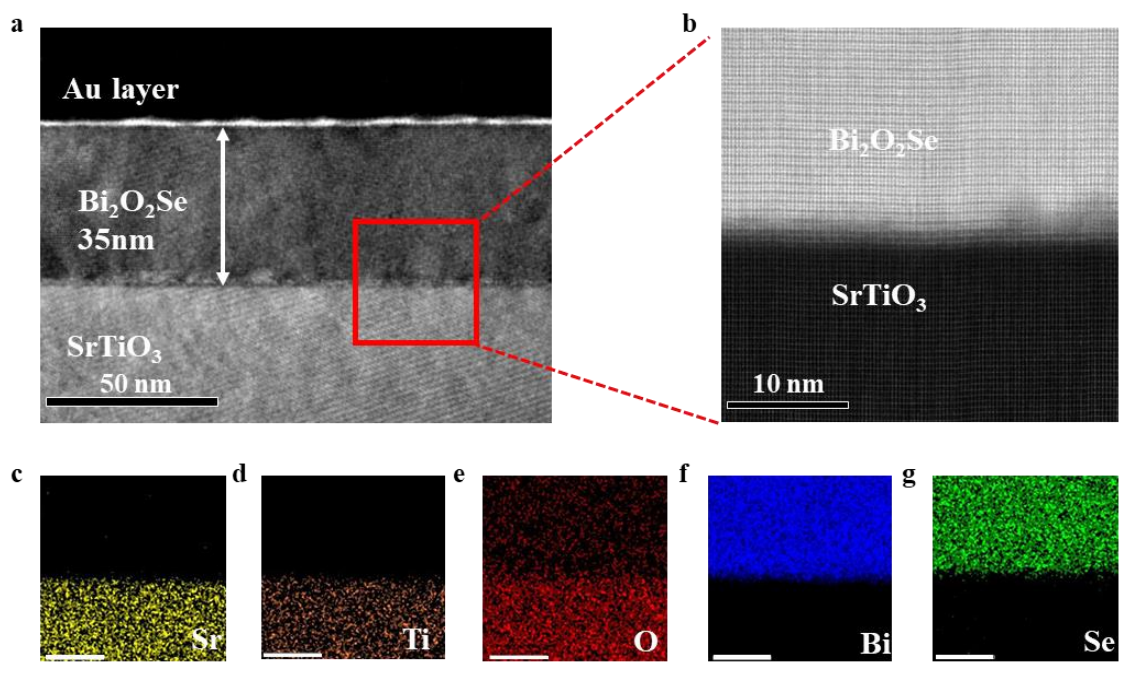

**Figure S1. (a)** Cross-sectional low-magnification TEM image of a 35nm-thick $Bi_2O_2Se$ film on the STO (001) substrate. **(b)** HAADF image of the interface region enlarged from the rectangular area in (a). **(c)-(f)** Corresponding EDX elemental maps of the $Bi_2O_2Se$/$SrTiO_3$ (001) interface for Sr, Ti, O, Bi, Se, indicating no obvious reaction layer located at the interface.

**Electrical measurements of Bi$_2$O$_2$Se thin films**

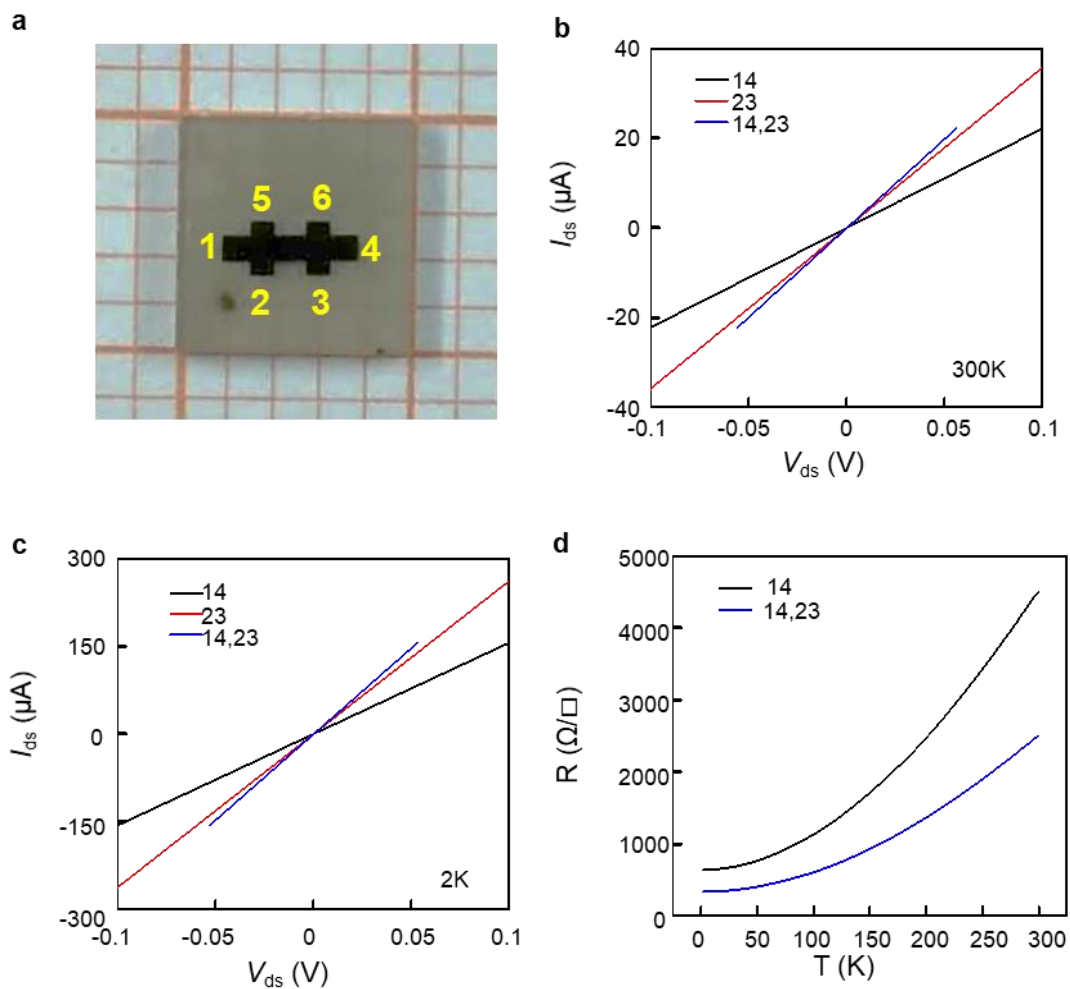

**Figure S2.** Ohmic contact tests of a 42nm-thick Bi$_2$O$_2$Se film. **(a)** Photograph of six-terminal shape Bi$_2$O$_2$Se film for electrical transport measurements. Metallic Indium was used to attach the current (1, 4) and voltage (2, 3, 5, 6) leads. **(b)-(c)** The Ohmic contact electrodes were tested through 2-probe and 4-probe I-V curves at 300K and 2K, respectively. Linear characteristics of the I-V curves indicates the Ohmic contacts were formed between Indium and Bi$_2$O$_2$Se films. **(d)** Temperature-dependent 2-probe and 4-probe resistance of Bi$_2$O$_2$Se thin film. The metallic-like conducting behavior of 2-probe R-T curve demonstrates the good contact formed between electrode and films.

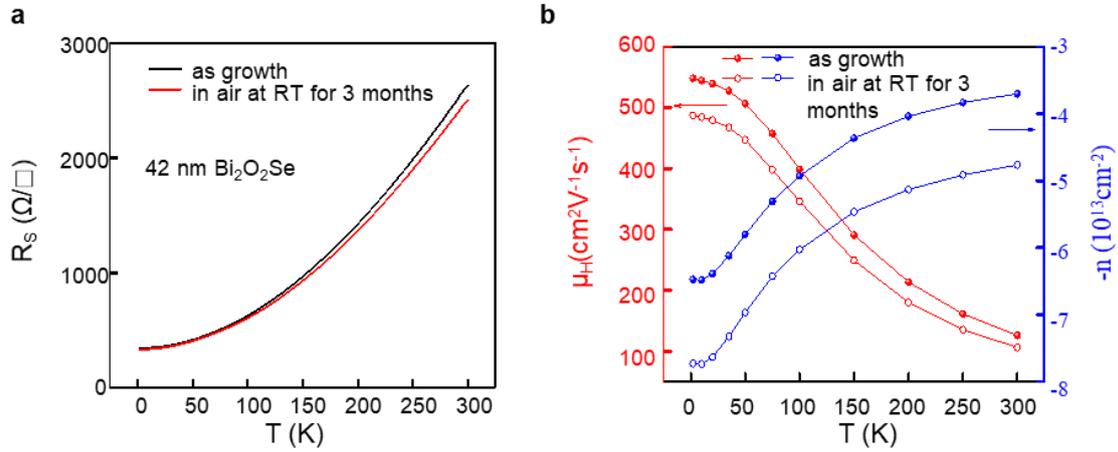

**Figure S3.** Electrical properties measurements and air-stability test of a $Bi_2O_2Se$ film with 42 nm thickness. **(a)** $R_s$-T curves of the as-grown sample and that exposed to air for ~3 months. The two $R_s$-T curves almost coincide with each other. **(b)** Temperature dependent Hall mobility and carrier concentration of the as-grown sample and that exposed to air for ~3 months. As the temperature decreases, the hall mobility increases which demonstrates phonon scattering is the main mechanism in the PLD-grown $Bi_2O_2Se$ films. The exposure to atmosphere for ~3 months leads to a slight increase in the electron concentration of $Bi_2O_2Se$ films. The exact mechanism behind the change in carrier density in $Bi_2O_2Se$ films is not well known, which is thought to be due to adsorption of other gasses on the film surface. The little changes in sheet resistance and electron carrier density result in the slight decrease in Hall mobility $\mu_H$. Above results demonstrate that the PLD-grown $Bi_2O_2Se$ films with good air-stability have great potential applications in electronics.